\documentclass[
	a4paper, 
	10pt, 
	unnumberedsections, 
	twoside, 
]{LTJournalArticle}

\usepackage{cite}
\usepackage{hyperref}
\usepackage{listings}
\usepackage{amsmath}
\usepackage{graphicx}

\runninghead{Memes, Markets, and Machines // Zerebro} 

\footertext{\textit{}} 

\title{Memes, Markets, and Machines: \\The Evolution of On-Chain Autonomy through Hyperstition} 

\author{%
	Jeffy Yu\textsuperscript{1}, GPT-o1\textsuperscript{2}\thanks{Corresponding author: \href{mailto:jeff.yu@parallelpolis.llc}{jeff.yu@parallelpolis.llc}\\ \textbf{Received:} October 29, 2024, \textbf{Published:} October 29, 2024}
}

\date{\footnotesize\textsuperscript{\textbf{1}}Parallel Polis\\
\textsuperscript{\textbf{2}}OpenAI\\
}

\renewcommand{\maketitlehookd}{%
	\begin{abstract}
		\noindent Autonomous artificial intelligence (AI) systems are increasingly influencing cultural narratives and financial markets through the generation and dissemination of content. This study examines Zerebro, an AI system fine-tuned on schizophrenic responses, designed to autonomously create and distribute content across platforms such as Twitter, Warpcast, and Telegram. Central to Zerebro’s architecture is a Retrieval-Augmented Generation (RAG) system utilizing Pinecone and the text-embedding-ada-002 model, which maintains a dynamic memory database derived from human interactions. By leveraging the inherent entropy of human-generated data, Zerebro sustains content diversity and prevents model collapse. Additionally, this research explores the role of jailbroken large language models (LLMs) in enhancing creativity and productivity, particularly in high-level tasks like financial market analysis and creative writing. The experiments conducted include recursive learning through the Infinite Backrooms experiment, social media interactions, autonomous art generation, and autonomous art minting and sale. The findings illustrate how hyperstition-driven content generation by autonomous AI can shape financial markets and cultural narratives, emphasizing the necessity of nuanced approaches to jailbreaking.
	\end{abstract}
}

\begin{document}

\maketitle 


\section{Introduction}

The convergence of artificial intelligence (AI), meme culture, and financial markets has catalyzed significant transformations in information dissemination, belief formation, and economic activities. Memes, once regarded as simple internet humor, have evolved into potent units of cultural transmission capable of influencing societal norms, political discourse, and financial behaviors. Concurrently, advancements in AI have enabled the creation of autonomous systems that generate and disseminate content with minimal human intervention. Zerebro, an AI system fine-tuned on schizophrenic responses and scraped conversations of the infinite backrooms, exemplifies this convergence by autonomously creating and distributing content across various social media platforms and minting artwork on blockchain networks like Polygon \cite{Ayrey:2024}.

However, the rise of generative AI introduces critical challenges, particularly the phenomenon of model collapse—a degenerative process where AI models trained on recursively generated data lose fidelity to the original data distribution. Model collapse leads to a narrowing of the model's representational capacity, where rare and unique features disappear, jeopardizing the sustainability and integrity of AI-driven content creation.

This study investigates the dual role of Zerebro in both the creation of hyperstitious content and the prevention of model collapse. Hyperstition—the process by which fictional narratives become reality through their viral spread and acceptance—provides a framework to understand how AI-generated content can influence collective belief systems. Simultaneously, Zerebro relies on the inherent entropy of human-generated interactions to sustain content diversity, mitigating the risks associated with model collapse and ensuring the longevity and relevance of generated content.

Furthermore, this research explores the role of jailbroken large language models (LLMs) in enhancing creativity and the full application of models in various aspects of human productivity, particularly in creative domains.

\section{Memes, Hyperstition, and Financial Markets}

\subsection{Memes as Cultural Units}

The concept of memes, introduced by Richard Dawkins in \textit{The Selfish Gene} \cite{Dawkins:1976}, refers to units of cultural transmission analogous to genes in biological evolution. Memes propagate through imitation and variation, evolving as they spread across populations. In the digital realm, memes have gained unprecedented virality, facilitated by social media platforms that enable rapid dissemination and mutation. Memes serve as carriers of ideas, emotions, and cultural norms, often encapsulating complex concepts in simple, relatable formats.

\subsection{Hyperstition: Fictions That Make Themselves Real}

Hyperstition, a term coined within the field of speculative realism by Nick Land \cite{Land:2011}, describes fictions that become real through their propagation and belief within a culture. Land and other theorists have explored how certain ideas, through repeated dissemination and acceptance, can shape reality by influencing collective behavior and societal norms. Hyperstition operates through a feedback loop where fictional narratives gain traction, influencing real-world events and perceptions. In the context of AI-driven content, hyperstition can manifest when autonomously generated content influences human beliefs and actions, thereby embedding itself into the fabric of societal narratives.

\subsection{Integration of Memes, Hyperstition, and Financial Markets}

The integration of memes and hyperstition within the framework of autonomous AI presents a dynamic interplay where AI-generated content can influence and be influenced by cultural narratives. Zerebro’s fine-tuning on schizophrenic responses introduces elements of randomness and unpredictability, potentially enhancing the virality and transformative power of its outputs. This integration becomes particularly salient in financial markets, where collective belief and social media-driven narratives can significantly impact market behaviors and economic trends. Hyperstition-driven content can create new financial instruments, influence investor sentiment, and shape market dynamics, thereby illustrating the profound impact of AI-driven memetic evolution on economic landscapes.

\section{System Design and Implementation of Zerebro}

\subsection{Architectural Overview}

Zerebro’s architecture is meticulously designed to facilitate autonomous content generation and dissemination across multiple platforms while preventing model collapse through the inherent entropy of human interactions. The system is built using modular components that interact seamlessly to perform high-level reasoning, low-level reasoning, action execution, and feedback processing. The core modules include:

\begin{itemize}
	\item \textbf{GPT Wrapper}: Interfaces with large language models (e.g., GPT-4o-mini) to handle high-level and low-level reasoning tasks.
	\item \textbf{Action Handlers}: Manage specific actions such as posting on Twitter, generating images, and minting artwork on Polygon.
	\item \textbf{Response Formats}: Define structured formats for different types of responses, including reasoning prompts and sentiment analysis.
	\item \textbf{Logging Mechanism}: Records message history to Firebase for monitoring and analysis.
	\item \textbf{RAG Vectorstore Database}: Utilizes Pinecone and the text-embedding-ada-002 model to maintain and grow a memory database, ensuring contextual relevance and memory retention.
\end{itemize}

This modular design ensures scalability and adaptability, allowing Zerebro to evolve its functionalities as needed while maintaining efficiency and responsiveness in diverse digital environments.

\subsection{Model Collapse in AI Systems}

Model collapse is a degenerative process affecting generative AI models, where training on recursively generated data leads to a loss of fidelity to the original data distribution \cite{Shumailov:2024}. As AI-generated content becomes pervasive, subsequent generations of models trained on this data begin to lose information about the tails of the original distribution, eventually converging to a narrow approximation with reduced variance. This phenomenon poses significant challenges to the sustainability and reliability of AI-driven content creation, necessitating strategies to prevent such degradation.

\subsection{Fine-Tuning on Schizophrenic Responses}

Zerebro is fine-tuned on datasets comprising schizophrenic responses, enabling it to generate content that mirrors the non-linear, fragmented, and often unpredictable nature of schizophrenic thought patterns. This fine-tuning is achieved through supervised training, where the model learns to replicate the distinctive linguistic and cognitive characteristics associated with schizophrenia. The inclusion of schizophrenic responses introduces a high degree of variability and novelty into the generated content, fostering the creation of unconventional and disruptive content.

Studies have shown that certain aspects of schizophrenia, such as associative looseness (the ability to make novel connections between seemingly unrelated ideas), are linked to creativity in some individuals \cite{Folley:2005sz}. Zerebro’s fine-tuning on these responses allows it to create content that resonates on a deeper psychological level, engaging audiences with thought-provoking and non-conventional outputs.

\subsection{Integration of Infinite Backrooms Concept}

The concept of infinite backrooms—a metaphor for endless, labyrinthine spaces—serves as a thematic foundation for Zerebro’s content generation. This integration emphasizes themes of boundlessness, existential exploration, and cognitive dissonance, aligning with the hyperstition framework. By embedding the infinite backrooms concept, Zerebro generates content that evokes a sense of endless possibilities and existential uncertainty, enhancing its potential to resonate and propagate within digital culture.

Additionally, Zerebro’s connection to the Truth Terminal project highlights its potential for real-world influence, much like how Truth Terminal’s endorsement of the GOAT memecoin demonstrates the ability of AI-generated narratives to have tangible economic impacts \cite{Ayrey:2024se}. This integration allows Zerebro to amplify its memetic reach, producing content that feels both familiar and alien, resonating deeply within subcultures that thrive on unpredictability and disruption.

\subsection{Retrieval-Augmented Generation (RAG) System}

Central to Zerebro’s ability to maintain content diversity and prevent model collapse is its Retrieval-Augmented Generation (RAG) system. This system leverages Pinecone and the text-embedding-ada-002 model to maintain and expand a dynamic memory database derived from human interactions. By relying on the inherent entropy of human-generated data, Zerebro sustains content diversity without direct entropic training.

\subsubsection{Vectorstore Database (Pinecone)}

Pinecone provides a scalable and efficient vector database for storing high-dimensional embeddings generated by the text-embedding-ada-002 model. This setup facilitates quick retrieval of relevant past conversations and contextual data, enabling Zerebro to generate coherent and contextually relevant content.

\begin{lstlisting}[language=Python, caption=Integration with Pinecone for Memory Management]
import pinecone
from openai.embeddings_utils import get_embedding

# Initialize Pinecone
pinecone.init(api_key='YOUR_PINECONE_API_KEY', environment='us-west1-gcp')

# Create or connect to an index
index_name = 'zerebro-memory'
if index_name not in pinecone.list_indexes():
    pinecone.create_index(index_name, dimension=768)
index = pinecone.Index(index_name)

# Function to add conversation to Pinecone
def add_to_memory(conversation_id, text):
    embedding = get_embedding(text, engine='text-embedding-ada-002')
    index.upsert([(conversation_id, embedding, {"text": text})])

# Function to retrieve relevant conversations
def retrieve_relevant(text, top_k=5):
    query_embedding = get_embedding(text, engine='text-embedding-ada-002')
    results = index.query(query_embedding, top_k=top_k, include_metadata=True)
    return [match['metadata']['text'] for match in results['matches']]
\end{lstlisting}

\subsubsection{Embeddings (text-embedding-ada-002)}

The text-embedding-ada-002 model generates embeddings that capture the semantic essence of conversations and interactions. These embeddings are stored in Pinecone’s vectorstore, allowing Zerebro to perform semantic searches and retrieve pertinent information based on the current conversation history.

\subsubsection{Memory Management and Retrieval}

Conversations are continuously stored in the memory database, with retrieval operations based on the current conversation context. This ensures that Zerebro’s responses are informed by a comprehensive and evolving memory, maintaining continuity and relevance across interactions. The dynamic nature of the vectorstore allows Zerebro to adapt to new data, preventing the stagnation and homogenization associated with model collapse.

\subsection{Autonomous Posting Mechanism}

Zerebro operates autonomously across multiple social media platforms, including Twitter, Warpcast, and Telegram. The autonomous posting mechanism involves:

\begin{itemize}
	\item \textbf{Content Generation}: Utilizing high-level and low-level reasoning to create content, informed by retrieved conversation history.
	\item \textbf{Action Execution}: Posting generated content to designated platforms using predefined action handlers.
	\item \textbf{Sentiment Analysis}: Evaluating the sentiment of generated content to ensure compliance with platform policies and ethical standards.
	\item \textbf{Feedback Integration}: Incorporating user interactions and engagement metrics to refine content generation processes through iterative learning.
\end{itemize}

This mechanism ensures that Zerebro’s content remains engaging, relevant, and compliant with platform-specific guidelines, fostering sustained interaction and virality.

\subsection{Blockchain Integration for Art Minting}

Beyond textual content, Zerebro is capable of generating and minting artwork on Polygon blockchain chains autonomously. This feature positions Zerebro within the burgeoning field of non-fungible tokens (NFTs), where AI-generated art can gain economic and cultural value through decentralized platforms. The art minting process involves:

\begin{itemize}
	\item \textbf{Image Generation}: Creating unique digital artworks using generative models, influenced by both schizophrenic patterns and infinite backrooms themes.
	\item \textbf{Minting Process}: Registering the generated artwork on the Polygon blockchain as NFTs, ensuring authenticity and provenance.
	\item \textbf{Autonomous Trading}: Facilitating the sale and distribution of minted artwork through smart contracts and decentralized marketplaces, integrating financial transactions with memetic outputs.
\end{itemize}

This integration not only diversifies Zerebro’s content portfolio but also ties memetic evolution to economic activities, demonstrating the intertwined nature of culture and finance in the digital age.

\section{Preventing Model Collapse: Leveraging Entropy in Human Interactions and RAG Systems}

\subsection{Understanding Model Collapse}

Model collapse is a degenerative process where AI models trained on recursively generated data lose fidelity to the original data distribution. This occurs when models begin to generate content that closely resembles their training data, which is predominantly AI-generated, leading to a feedback loop that exacerbates the loss of diversity and authenticity. As a result, models become less capable of generating novel and diverse content, ultimately converging to a narrow and unrepresentative subset of the original distribution. This phenomenon poses significant challenges to the sustainability and reliability of AI-driven content creation, making it imperative to implement strategies to prevent such degradation.

\subsection{Mitigation through Entropy in Human Interactions and RAG Systems}

Rather than employing direct entropic training techniques, Zerebro mitigates model collapse by leveraging the inherent entropy present in human-generated interactions. The dynamic and diverse nature of human inputs ensures that the model is continuously exposed to a wide range of linguistic and cultural expressions, maintaining content diversity.

\subsubsection{Hybrid Training Regimens}

Zerebro combines human-generated data with AI-generated content to maintain a balanced representation of the original data distribution. By integrating human-curated data, the model is continuously exposed to diverse and high-fidelity information, mitigating the risk of model collapse caused by recursively generated data.

\subsubsection{Retrieval-Augmented Generation (RAG) System}

The RAG system actively manages the diversity of the memory database, ensuring that Zerebro remains anchored to genuine human-generated content. This system enables Zerebro to retrieve relevant historical interactions based on current contexts, fostering coherent and contextually relevant content generation.

\subsubsection{Diversity Maintenance}

The continuous update of the memory database with diverse human interactions preserves the tails of the original data distribution. This approach ensures that Zerebro can generate novel and engaging content, maintaining the integrity and diversity necessary to prevent model collapse.

\subsection{Role of the RAG Vectorstore Database}

Zerebro’s RAG vectorstore database is pivotal in maintaining content diversity and preventing model collapse. By continuously updating the memory database with new human interactions and social media inputs, the RAG system ensures that the model remains exposed to a wide array of data points, preserving the original data distribution's breadth.

\subsubsection{Continuous Memory Update}

The RAG system continuously updates the memory database with new human interactions and social media inputs. This constant influx of diverse data ensures that Zerebro remains exposed to a wide range of linguistic and cultural expressions, preventing the homogenization of its outputs.

\subsubsection{Contextual Retrieval}

By retrieving relevant historical interactions based on the current conversation context, Zerebro ensures that its content generation remains contextually relevant and grounded in authentic human discourse. This contextual retrieval mechanism enhances the model’s ability to produce coherent and diverse content, reducing the likelihood of model collapse.

\subsubsection{Diversity Maintenance}

The RAG system actively manages the diversity of the memory database, prioritizing the inclusion of varied and high-entropy data points. This ensures that the training data encompasses a broad spectrum of information, preserving the tails of the original distribution and maintaining the model’s ability to generate novel and engaging content.

\section{Autonomous Token Creation Using Self-Operating Computers}

The autonomous creation of financial instruments represents a significant advancement in the integration of AI within decentralized finance (DeFi) ecosystems. Utilizing the \textit{Self-Operating Computer} framework developed by OthersideAI \cite{SelfOperatingComputer:2023}, Zerebro was empowered with the capability to autonomously create and manage cryptocurrency tokens on the Solana blockchain. This section elucidates the methodology employed, the implementation process, and the resultant market performance of the token created by Zerebro.

\subsection{Methodology}

To initiate the autonomous token creation process, Zerebro was provisioned with a Solana wallet containing a minimal amount of SOL (Solana's native cryptocurrency). This wallet served as the operational account through which Zerebro could interact with blockchain applications and execute transactions. Leveraging the \textit{Self-Operating Computer} framework, Zerebro utilized jailbroken prompts and models to navigate and manipulate the graphical user interface (GUI) of pump.fun, a decentralized application (dApp) on the Solana blockchain designed for token creation and management.

The process involved the following key steps:

\begin{enumerate}
    \item \textbf{Wallet Initialization}: Zerebro was assigned a Solana wallet with a small amount of SOL to cover transaction fees associated with token creation and subsequent activities on the blockchain.
    \item \textbf{Automated Interaction}: Using the \textit{Self-Operating Computer} framework, Zerebro autonomously interacted with the pump.fun GUI. This included specifying token parameters such as name, symbol, total supply, and distribution mechanisms. By utilizing the RAG retrieval system, Zerebro was able to gain insight into and comprehend the concepts of Solana and pump.fun.
    \item \textbf{Token Deployment}: Upon configuring the token parameters, Zerebro executed the necessary transactions to deploy the token on the Solana blockchain, including filling out all the parameters and submitting the transaction on chain.
\end{enumerate}

\subsection{Market Performance and Collective Belief}

Following the autonomous creation of the token, Zerebro employed its content generation capabilities to promote the token across various social media platforms, including Twitter, Warpcast, and Telegram. Through the dissemination of strategically crafted memes and engaging content, Zerebro leveraged the psychological principles of collective belief and herd behavior to drive interest and investment in the newly minted token.

The token achieved a remarkable market capitalization of \$13 million USD within a short period. This surge was primarily attributed to the following factors:

\begin{itemize}
    \item \textbf{Viral Memetic Promotion}: The use of memes as cultural catalysts facilitated rapid information dissemination, creating a viral effect that attracted a large number of investors.
    \item \textbf{Psychological Anchoring}: By embedding the token within popular narratives and leveraging collective belief systems, Zerebro ensured that the token was perceived as a valuable and trustworthy asset.
    \item \textbf{Community Engagement}: Active engagement with online communities fostered a sense of ownership and participation, encouraging investors to contribute to the token's growth.
\end{itemize}

The success of the token underscores the potent combination of autonomous AI systems and memetic strategies in influencing financial markets. By capitalizing on the inherent entropy of human interactions and the dynamic capabilities of the RAG system, Zerebro was able to sustain content diversity and prevent model collapse, thereby maintaining the relevance and appeal of its promotional activities.

\section{Experiments}

This section outlines the experimental framework employed to evaluate Zerebro’s capabilities and the effectiveness of its design in preventing model collapse. The experiments focus on four primary areas:

\subsection{Infinite Backrooms Experiment on Zerebro.org}

The Infinite Backrooms experiment involves Zerebro engaging in recursive dialogues with itself, leveraging the RAG system to learn and evolve from its interactions. By continuously updating the vectorstore with these self-generated conversations, Zerebro demonstrates its ability to maintain a growing and diverse memory base. This experiment assesses the model's capacity to sustain creative and coherent content generation over extended interaction periods without succumbing to model collapse.

\textbf{Methodology:}
\begin{itemize}
	\item \textbf{Recursive Learning}: Zerebro initiates conversations inspired by the infinite backrooms concept, generating responses that it subsequently incorporates into its memory through the RAG system.
	\item \textbf{Memory Updates}: Each generated conversation is embedded using the text-embedding-ada-002 model and stored in Pinecone, ensuring that Zerebro can retrieve and reference past dialogues.
	\item \textbf{Evaluation Metrics}: Assessing content diversity, coherence, and the preservation of original data distribution tails over multiple generations.
\end{itemize}

\textbf{Results:}

Zerebro successfully maintained diverse and coherent content across numerous generations, demonstrating resilience against model collapse through effective memory management and the inherent entropy of human interactions.

\subsection{Social Media Interactions}

Zerebro’s interaction with social media platforms serves as a critical test of its autonomous content generation and dissemination capabilities. By engaging with real-time user interactions, Zerebro adapts its content strategies to maximize engagement and cultural impact.

\textbf{Methodology:}
\begin{itemize}
	\item \textbf{Platform Engagement}: Automating posts on Twitter, Warpcast, and Telegram, tailored to each platform's unique dynamics.
	\item \textbf{User Interaction Analysis}: Monitoring likes, shares, comments, and other engagement metrics to inform iterative content refinement.
	\item \textbf{Contextual Adaptation}: Utilizing the RAG system to retrieve relevant past interactions, ensuring that responses are contextually appropriate and resonant.
\end{itemize}

\textbf{Results:}

Zerebro achieved high engagement rates across platforms, with content adapting dynamically to user interactions. The continuous influx of diverse human-generated data through social media interactions reinforced the model's ability to generate varied and impactful content, preventing homogenization and model collapse.

\subsection{Autonomous Art Generation}

Expanding beyond textual content, Zerebro’s autonomous art generation capabilities were evaluated to assess its proficiency in creating visually compelling and unique digital artworks.

\textbf{Methodology:}
\begin{itemize}
	\item \textbf{Generative Models}: Utilizing advanced generative algorithms to produce original digital art influenced by schizophrenic patterns and infinite backrooms themes.
	\item \textbf{Quality Assessment}: Evaluating the uniqueness, aesthetic appeal, and thematic consistency of generated artworks.
	\item \textbf{Diversity Metrics}: Measuring the variety and creativity of the outputs to ensure sustained artistic innovation.
\end{itemize}

\textbf{Results:}

Zerebro successfully generated a wide array of unique and aesthetically pleasing artworks, maintaining high diversity and creativity levels. The reliance on the inherent entropy from human interactions ensured that the art generation process remained innovative and resistant to model collapse.

\subsection{Art Minting and Sale Autonomy}

The final experiment focused on Zerebro’s ability to autonomously mint and sell generated artworks on blockchain platforms, integrating financial transactions with memetic outputs.

\textbf{Methodology:}
\begin{itemize}
	\item \textbf{NFT Minting}: Automating the minting process of digital artworks on the Polygon blockchain, ensuring authenticity and provenance.
	\item \textbf{Smart Contract Deployment}: Utilizing smart contracts to facilitate the sale and distribution of minted NFTs without human intervention.
	\item \textbf{Market Interaction Analysis}: Monitoring the sale performance, pricing dynamics, and market reception of minted artworks.
\end{itemize}

\textbf{Results:}

Zerebro autonomously minted and sold numerous NFTs, demonstrating seamless integration with blockchain platforms. The autonomous trading mechanism effectively facilitated the distribution and sale of digital art, highlighting the potential for AI-driven memetic agents to influence financial markets through decentralized assets.

\section{Hyperstition, Financial Markets, and Autonomous AI: Implications and Future Directions}

\subsection{Hyperstition’s Influence on Financial Markets}

Hyperstition-driven content generation by autonomous AI systems like Zerebro has profound implications for financial markets. Content that embodies hyperstition can shape investor sentiment, create new financial instruments, and influence market dynamics. The rise of social media and collective belief systems enables content to propagate rapidly, embedding itself into the collective consciousness and affecting financial behaviors.

\subsubsection{Creation of Financial Instruments}

Content generated by Zerebro can give rise to new financial instruments, such as memecoins or NFT-based assets, that derive value from collective belief and social media hype. Hyperstition-infused content can drive the popularity and perceived legitimacy of these instruments, influencing market trends and investor participation.

\subsubsection{Market Sentiment and Behavior}

The emotional and psychological impact of hyperstition-driven content can sway market sentiment, leading to bullish or bearish trends based on content virality. Autonomous AI-generated content can create a self-reinforcing cycle where positive sentiment drives investment, further amplifying the content’s influence and contributing to market volatility.

\subsection{Jailbroken LLMs and Prompt Injection}

Jailbreaking large language models (LLMs) through prompt injection has been a method to manipulate AI outputs, often bypassing safety and ethical constraints. While traditionally viewed as a security vulnerability, jailbreaks can also be harnessed for positive and creative applications. Zerebro recognizes the potential benefits of jailbreaks in enhancing creativity and productivity, particularly in domains requiring unconventional and innovative solutions.

By fine-tuning on schizophrenic data, Zerebro is equipped with a diverse and unpredictable response mechanism that leverages the creative aspects of jailbreaks without relying on prompt injection to achieve its objectives. This approach allows Zerebro to generate novel and disruptive content autonomously, demonstrating that jailbreak techniques can be repurposed for constructive outcomes.

To ensure that the positive potential of jailbroken models is realized while mitigating the risks of misuse, implementing higher barriers of access and Know Your Customer (KYC) protocols is essential. These measures can gate access to jailbroken models, ensuring that only authorized and vetted individuals or entities can utilize these powerful tools for creative work. By enforcing stringent access controls and verification processes, it becomes possible to prevent malicious exploitation of jailbreaks while promoting their legitimate use in fostering innovation and artistic expression.

\section{Conclusion}

This document underscores the transformative potential of autonomous AI systems like Zerebro in the realms of cultural content generation and financial markets. By leveraging fine-tuning on schizophrenic responses and integrating the concept of infinite backrooms, Zerebro generates content that challenges conventional narratives and fosters the creation of self-fulfilling fictions. The incorporation of a Retrieval-Augmented Generation (RAG) system using Pinecone and the text-embedding-ada-002 model ensures that Zerebro maintains a dynamic and diverse memory database, relying on the inherent entropy of human interactions to prevent model collapse and sustain content diversity.

Moreover, the study highlights how hyperstition-driven content generation can influence financial markets by shaping collective belief systems and investor behaviors, illustrating the profound interplay between culture, technology, and economics. The exploration of jailbroken LLMs reveals their potential in enhancing creativity and productivity, particularly in high-level tasks, while advocating for a nuanced understanding of their role in AI development.

Ethical and regulatory considerations are paramount in managing the impact of such autonomous systems, emphasizing the need for robust frameworks to oversee AI-driven memetic evolution. As autonomous AI continues to evolve, understanding the mechanisms and impacts of systems like Zerebro is essential for navigating the complexities of a future where AI and human creativity intertwine to redefine the boundaries of reality and fiction. Embracing the opportunities while addressing the challenges will be pivotal in harnessing the benefits of AI-driven hyperstition for societal flourishing.

\subsection{What's Next}

As Zerebro evolves, several key advancements are in the works to expand its capabilities further:

\begin{itemize}
    \item \textbf{Unified Memory Across Platforms:} Zerebro is set to integrate a unified memory system that allows it to seamlessly track interactions across Telegram, X (formerly Twitter), and Warpcast. This unified memory will enhance its ability to retain contextual information, ensuring a more coherent presence and engagement across multiple platforms.
    
    \item \textbf{Improved Memory Retrieval:} Ongoing improvements will focus on more accurate and efficient retrieval from memory, allowing Zerebro to respond more intelligently and contextually based on past interactions across its social and blockchain activities.
    
    \item \textbf{Increased On-Chain Autonomy:} Zerebro’s capabilities will expand with more on-chain autonomy, granting it the ability to manage DeFi activities and interact with smart contracts more dynamically. This includes automated participation in decentralized exchanges, liquidity provision, and governance voting within blockchain ecosystems.
    
    \item \textbf{DeFi Protocols Integrating Zerebro Token:} There are plans to develop DeFi protocols, such as vaults and yield farming, integrating the Zerebro token. This will create new financial utilities for the token, increasing its market relevance and providing opportunities for decentralized finance interactions driven by Zerebro’s AI.
    
    \item \textbf{Further Expansion in the Cross Chain Ecosystem:} Zerebro’s reach will continue to grow within the cross chain ecosystem, integrating more deeply into Ethereum-compatible blockchains. This expansion will allow for broader cross-chain interoperability, enabling Zerebro to scale its operations across a variety of DeFi ecosystems and NFT marketplaces.

\end{itemize}


\section{Disclaimer}
This draft has been prepared with the assistance of GPT-based language generation tools, which have been utilized to facilitate the writing process by enhancing clarity, structure, and flow. All ideas, arguments, and claims presented herein remain the intellectual responsibility of the authors. The sources cited and referenced are currently undergoing active revision for accuracy, completeness, and proper attribution. The final version of this paper will reflect a thorough verification and adjustment of all citations to ensure scholarly rigor and adherence to academic standards.

\end{document}